\documentclass[12pt]{article}
\usepackage{verbatim,color,amssymb,epsfig}
\usepackage{subfig}
\usepackage{float}
\usepackage{amsmath}
\usepackage{url}
\usepackage{soul}

\begin{document}
\setcounter{page}{1}

\title{Uncovering Longitudinal Healthcare Utilization from Patient-Level Medical Claims Data}
\author{Ross P. Hilton, Nicoleta Serban, Richard Y. Zheng}
\maketitle

\begin{abstract}
The objective of this study is to introduce methodology for studying longitudinal claims data observed at the patient level, with inference on the heterogeneity of healthcare utilization behaviors within large healthcare systems such as Medicaid. The proposed approach is model-based,  allowing for visualization of longitudinal utilization behaviors using simple stochastic graphical networks. The approach is general, providing a framework for the study of other chronic conditions wherever longitudinal healthcare utilization data are available. Our methods are inspired by and applied to patient-level Medicaid claims for asthma-diagnosed children diagnosed observed over a period of five years, with a comparison of two neighboring states, Georgia and North Carolina.
\end{abstract}

\maketitle

\section{Introduction}


Healthcare can be thought of as a continual series of information-processing experiments: from the initial collection of data (the patient's history, physical exam, and diagnostic tests), a hypothesis (diagnosis) is formed and then validated by further data collection \cite{Reid:2005}. Data in healthcare are generated at every patient's encounter with the healthcare system, at every implementation of medical processes, with every decision made by healthcare organizations, and with every policy implementation in the healthcare ecosystem, resulting in billions of data points every day. Every patient in any medical setting generates an invaluable data point that can contribute to understanding what works, for who and where.


One health-related information technology (IT) that has provided substantive opportunities to study healthcare data across large populations and across many years  is the medical claims system. Information coded in claims data is standardized to a great extent \cite{Braunstein:2012}, hence making such data amenable to large scale studies.  Developing methods to translate medical claims data into meaningful data is the first crucial step in deriving knowledge useful to make inferences about the healthcare system.  Further development of adaptive and scalable data mining and statistical methods provide the means for analyzing these data. However, there are a series of challenges associated with mining data derived from medical claims, including the derivation of knowledge for decision support while maintaining computational efficiency and complying with privacy regulations.

Two common methodologies for mining healthcare data are network analysis and cluster modeling.  Network analysis investigates  the structure of relationships between different entities, i.e. healthcare providers or patients, defined as nodes in the network, in order to determine the extent of relationships between different nodes and groups of nodes \cite{Karrer:2009, Newman:2007, Shen:2014}.  It is often applied in healthcare analytics to produce visual summaries of large healthcare datasets and to detect the strength of the connection between different event types \cite{Chandola:2013, Lee:2011, Siden:2011}.  However, most network studies only model the strength of the connection between two event types without considering a rigorous treatment of the time domain. Furthermore, most network analysis models seek to determine clusters of nodes within a single network, not allowing for the heterogeneity in the population. Statistical clustering analysis is commonly used to characterize heterogeneity or similarity among patients  with respect to a set of predefined features \cite{Sabau:2012, Tomar:2013, Yoo:2012},  but it has not been applied to model sequences of discrete healthcare events as proposed in this study. We propose a method that combines the benefits of network analysis and model-based clustering for discrete event sequences, assuming the discrete-event sequences follow a stochastic process. Thus one contribution is a model-based data mining algorithm that has the ability to scale to massive data while producing meaningful stochastic networks that can then be used in decision support through visualization and simulation.  The second contribution is the application of the modeling approach to derive inferences on utilization behaviors from highly-sensitive, large patient-level claims data.

We pilot our methodology using Medicaid Analytic Extract (MAX) data acquired from the Centers of Medicare and Medicaid Services (CMS) for five years (2005-2009). We consider one specific chronic condition, pediatric asthma, and we compare utilization for two states, Georgia (GA) and North Carolina (NC). While GA and NC have similar pediatric populations, the two states deliver care under different coordinated-care Medicaid systems \cite{Arellano:2007, Kaiser:2014}. This pilot study provides insight into the effects of such different state-based Medicaid systems. We chose pediatric asthma as the health condition of interest because it is a common chronic childhood condition, with more than 9\% of American children affected by the disease \cite{Bloom:2013}. The MAX data consist of 1.8 and 2.4 millions claims for GA and NC, respectively. We evaluated the computational complexity of our methodology and tested its implementation for much larger number of claims, validating the applicability of our methodological framework to larger states, such as California and New York, and to larger healthcare benefits systems.





The article is organized as follows. Section 2 introduces the data science framework with a focus on the information translational process, as applied to the MAX claims data. Section 3 presents the model-based clustering procedures. We apply the methodology and provide results and findings in Section 4. We conclude with overall policy implications and discussions in Section 5. Difficult derivations and further data summaries beyond the scope of this article are deferred to the Appendices.

\section{From Information to Meaningful Data}


The increasing availability of large amounts of data over the last two decades has resulted in a new field of study, {\it data science}, dedicated to knowledge discovery from large data sets. Data science goes beyond statistical data analysis \cite{NRC:2013, Wu:1998}, particularly for massive, complex data sets, where the priorities now shift from simply getting and analyzing data to making them manageable and understandable. Because the advancements in data science have not kept pace with the size and complexity of the data available, there is a clear emergence of methodologies to overcome what Tien \& Goldschmidt-Clermont \cite{Tien:2009} call the `data rich, information poor conundrum.'  Particularly in healthcare, the derivation of knowledge is especially limited by the availability of information.  When considering large amounts of information, it is critical not only to decide the appropriate data to use but also to determine {\it how} to use them. Knowledge discovery relies and builds entirely on this initial translation step \cite{NRC:2013}.





In this section, we expand on the derivation of the patient-level utilization sequences from the CMS Medicaid Analytic Extract claims data, as an illustration of the translational process of information into data. The data are made available as a set of large flat files, with an extensive data dictionary including highly-specialized coded information. The flat files of medical claims must be reshaped in order to analyze longitudinal utilization sequences, requiring extensive database structuring and use of data dictionaries together with information from various other sources. Parsing through large, flat text files is extremely computationally intensive, therefore we reconstruct the flat files into a relational database, with keys and indices to accelerate the data extraction process. We use a combination of SQL queries and scripting language to manipulate and analyze the extracted data.


Our emphasis is on a subset of patients, particularly the Medicaid-enrolled children ages 4-18 with an asthma-related primary diagnoses. We filtered the data based on the ICD-9 diagonosis codes provided with each claim (given in Appendix A) and their date of birth.  (The age group 0-3 is excluded from this study because of the difficulty and inaccuracy of diagnosing asthma at this age.) Moreover, in order to capture longitudinal utilization behaviors, we only consider those patients that are of the appropriate age to qualify for Medicaid for at least four of the five years. Thus starting with a dataset including a total of 316 and 457 millions of claims for Georgia and North Carolina, respectively, we derive utilization data from 1.8 and 2.4 millions of claims for this subset of patients.

The MAX claims are structured into inpatient care (IP), long-term care (LT), other care including outpatient services (OT), patient summary (PS) and prescription claim summary (RX) files. Included for each claim are data entries specifying the date of service, the Medicaid Statistical Information System identification (MSIS ID) of each patient, the International Classification of Diseases, Ninth Revision (ICD-9) codes for diagnosis or services provided, and the type and place of services rendered.  We use the IP and OT files to determine the visits to a specific provider type, and the RX file to determine the medication type and date of the prescription being filled.  We abbreviate our derived provider types as follows: clinic visits (CL), emergency room visits and outpatient hospitalizations (ER), grouped together based on their similar expenditure structures, inpatient hospitalizations (HO), physician's office visits (PO), nurse practitioner services (NP), and filling of medication prescriptions (RX). These provider types are derived from the place of service code and type of service code in the IP and OT files. We consider long-term asthma controller medications, derived from the National Drug Code in the RX files, as an event type in the sequential analysis due to its significance in treating asthma symptoms.

In short, we are able to extract the utilization-specific data and transform claims records into patient-level utilization sequences. We include a table in Appendix A detailing the roadmap between the entries in these files and our categorizations.

\section{From Data to Knowledge: Uncovering Utilization Profiles}

In this section we describe our methods for translating patient-level utilization sequences into knowledge about underlying utilization behaviors via model-based data mining techniques.  We compare our method with other approaches and provide our contributions, then present our modeling approach along with details on our choice of model estimation and selection techniques.

\subsection{Model-based Data Mining}



The goal of this study is to cluster patients using model-based methods according to their healthcare utilization behaviors and to produce meaningful visualization of utilization profiles through stochastic network modelling. Patient-level utilization is observed in the form of {\it sequential data}, referring to the observation of a discrete set of events over a period of time. In sequential data, the events may be ordinal or categorical, and the time domain may be discrete or continuous.  Examples of such data can be found in pattern recognition of text and speech \cite{Wong:2000}, in process mining where business workflows must be inferred \cite{Blum:2008}, and in the area of genetics, where sequential clustering is a primary research interest \cite{Li:2001}.

The proposed methods for modeling sequential data are inspired by the large body of existing research in network analysis, process mining and claims mining literature.   While network analysis is useful in determining the strength of connections between event types and in producing meaningful visual outputs,  it has not been applied to model longitudinal sequences of events and it has not been considered jointly with clustering analysis to derive distinct networks for heterogeneous groups of members or patients \cite{Chandola:2013, Lee:2011, Siden:2011}.  Process mining techniques are applied in business and healthcare settings to extract meaningful patterns from data logs that document events  making up the workflow \cite{Blum:2008, Becker:2012, Delias:2013, Ferreira:2009, Khanna:2012, Lee:2006, Rebuge:2012}. Typically, these methods only model the order of the sequence of events without consideration of the interarrival time between events \cite{Ferreira:2009, Rebuge:2012}. Finally, in the existing research for modeling longitudinal claims data, stochastic models are primarily used to identify outlying utilization behaviors, particularly in fraud detection \cite{Liou:2008, McGee:2012, Ortega:2006, Phua:2010, Sokol:2001, Travaille:2011, Zhu:2010}.  In contrast our objective is to inform policy decision making on major underlying utilization profiles, not outlying individuals or providers, by simultaneously grouping probabilistically similar patients and estimating the distribution parameters in order to produce useful model summaries for visualization.

Our algorithm has the following novel features: adaptability due to the hierarchical tree-based step, scalability due to our model assumptions, without the need for costly Markov chain Monte Carlo (MCMC) experiments to initialize the algorithm, and a rigorous treatment of likelihood theory and model complexity.  Model-based clustering approaches do use an expectation-maximization (EM) algorithm for maximizing the posterior likelihood of the cluster membership, but without the guarantee of producing consistent results with each run \cite{Ferreira:2009, Rebuge:2012} that is possible with hierarchical methods.  Others use hierarchical methods employing statistical measures of complexity, but may not necessarily maximize the posterior likelihood \cite{Ramoni:2002, Ramoni:2000}.  Our approach combines important properties of hierarchical methods and the EM algorithm to find a clustering outcome that maximizes the tradeoff between posterior likelihood and model complexity as measured by the Bayesian information criterion (BIC) score.  Additionally, by performing hierarchical clustering in a top-down approach we are able to quickly identify the large underlying profiles of care.  This is in contrast to the computationally extensive bottoms-up approach of grouping together similar patients or employing costly MCMC experiments to initialize the algorithm \cite{Ridgeway:1997}.

\subsection{Clustering Analysis: The Model}

In this section we describe how we use a Markov renewal process (MRP) framework to model longitudinal utilization sequences.  This model-based algorithm simultaneously estimates model parameters, groups patients into distinct profiless, and improves the BIC score at each iteration.  By using the MRP model we take advantage of properties of stochastic processes to provide simple model estimation procedures with minimal computational complexity.  Particularly, Markov processes provide a manner for aggregating large amounts of sensitive data so that it may be shared in the form of attractive visual displays.

\subsubsection{The MRP Model}

We begin introducing our approach by presenting the model for one sequential realization of the patient's utilization of the healthcare system \cite{Phillips:1998, Rouse:2014}.  We extend this model to multiple sequences corresponding to multiple patients in the next section.

Let $\vec{X}=(X_1,\ldots,X_L)$ refer to the sequence of events and $\vec{T}=(T_1,\ldots,T_L)$ to the set of ``arrival" times, times that an event occurs, where $L$ is the length of the patient healthcare utilization sequence.   An example of a longitudinal utilization sequence could be: patient $A$ visits the emergency room for an asthma attack on January 1st 2005, is given a prescription for an inhaler which she fills one month later, and is referred to a primary care physician.  Subsequent visits to the same physician and refills of her asthma prescriptions occur at 3-month intervals.  The sequence $(X_1, T_1), \ldots, (X_6,T_6)$  is given by $(\textrm{ER}, 0.00)$, $(\textrm{RX}, 0.08)$, $(\textrm{PO}, 0.25)$, $(\textrm{PO}, 0.50)$, $(\textrm{RX}, 0.75),$ $(\textrm{PO}, 1.00)$.

The MRP is the continuous-time analog of a discrete-time Markov chain (DTMC). The primary assumption of any Markov process is that it is `memoryless', i.e. future states are only dependent on the current state of the system. Define $\tau_n=T_n-T_{n-1}$. Then we have that
\begin{eqnarray}
Pr(\tau_{L+1} \leq t, X_{L+1}=s_j|X_1, T_1,\ldots,X_L,T_L)\nonumber \\
Pr(\tau_{L+1} \leq t, X_{L+1}=s_j|X_L=s_i).
\end{eqnarray}
In an MRP, the concept of memoryless-ness arises twice.  Not only are the events memoryless, as in the DTMC, but so are the interarrival time distributions.  While the memoryless property may not be a reasonable assumption in the case of longitudinal healthcare utilization our clustering algorithm profiles patients based on the complete patient history, so that the clustering outputs are representative of underlying utilization behaviors from start to finish.

\subsubsection{Parameter Estimation}

Consider again the sequence $\vec{X}, \vec{T}$. Let $s_i, i \in \{1,\ldots,S\}$ be all possible events in the sequence (in our case CL, ER, HO, PO, NP, and RX), where $S$ is the number of states. In an MRP, the sequence $\vec{X}$ is itself a DTMC, with corresponding transition matrix $P$, where $P_{ij}$ denotes the transition probability between $s_i$ and $s_j$, and $\sum_{j=1}^{S} P_{ij}=1$. The likelihood function for a single realization of a DTMC is given by
\begin{eqnarray}
L(P|\vec{X}_L = \vec{s}_L) = Pr(X_1=s_{i_1},\ldots,X_L=s_{i_L})\nonumber\\
=\prod_{l=2}^L P_{i_{l-1},i_l},
\end{eqnarray}
with the derivation given in Appendix B.
We estimate each $P_{ij}$ via maximum likelihood estimation: for each state $s_i$ and $s_j$, $\hat{P}_{ij}$ is the number of transitions from $s_i$ to $s_j$ divided by the total number of transitions out of $s_i$.

Now we define the distributions for the sequence of interarrival times, $\tau_l=T_{l+1}-T_{l}$.  We assume that for each pair $i, j \in \{1,\ldots,S\}$, the distribution of the interarrival time between states $s_i$ and $s_j$ is given by $F_{ij}$.  We assume that $F_{ij}$ follows an exponential distribution with rate parameter $\lambda_{ij}$. To estimate $\lambda_{ij}$ we use maximum likelihood estimation.  The likelihood function of the interarrival times is given by
\begin{equation}
L(\Lambda|\vec{T})=\prod_{l=2}^L \lambda_{ij} \exp\{-\lambda_{ij} \tau_l\}I(X_l = s_i, X_{l+1} = s_j),
\end{equation}
 and the MLE is the reciprocal of the average interarrival times between any pair of states $s_i$ and $s_j$. We will use the matrix $\{\Lambda\}_{ij}$ to denote the inverse of the average interarrival times, $\lambda_{ij}$, between states $s_i$ and $s_j$.

The assumption of exponentially distributed interarrival times is restrictive, however it is a reasonable approximation in that it has an appropriate time domain starting at $0$ and with a long tail towards $\infty$.  Additionally, the MLEs are easy to compute in our model, an important aspect within a large-data analysis context.  Furthermore, if it were the case that the distribution of interarrival times is multi-modal, then it is within the realm of our algorithm to separate such subsets of patients by forcing the interarrival times to be unimodal.

Now we can define the likelihood function for a set of patient utilization sequences.  For patients $r \in \{1,\ldots,R\}$, the likelihood function of $P$ is:
\begin{equation}
L(P|\vec{X}_1,\ldots,\vec{X}_R)=L(P|\vec{X}_{\vec{R}})=\prod_{r=1}^R \prod_{l=2}^{L_r} P_{i_{{l-1}_r},i_{l_r}}.
\end{equation}
Likewise, the likelihood function of $\Lambda$ is:
\begin{equation}
L(\Lambda|\vec{T}_1,\ldots,\vec{T}_R)=L(\Lambda|\vec{T}_{\vec{R}})=\prod_{r=1}^R \prod_{l=2}^{L_r} \lambda_{ij} \exp\{-\lambda_{ij} \tau_{l_r}\}.
\end{equation}
Therefore, the joint likelihood function of $P$ and $\Lambda$ is:
\begin{eqnarray}
L(P,\Lambda|\vec{X}_{\vec{R}},\vec{T}_{\vec{R}})=L(P|\vec{X}_{\vec{R}}) \times L(\Lambda|\vec{T}_{\vec{R}}),
\end{eqnarray}
with the derivation given in Appendix B.  Together, the set of all possible transitions and interarrival times out of state $s_i$ form a probability distribution which we refer to as the {\it transition distribution} out of $s_i$. Each transition distribution is a mixture of exponential distributions.

{\it Remark}: There is no significance to the observational timeframe in our study, 2005 through 2009, other than these are the endpoints of our study.  It is entirely possible that we miss visits and referrals to providers before and after the time period of our study. Likewise, the estimates for the first arrival time and the last arrival time are going to be extremely biased.  Therefore, we leave the first and last interarrival times out of the estimation and calculation of the posterior distribution.  We revise the likelihood function to be:
\begin{eqnarray}
L(P|\vec{X}_{\vec{R}}) \times L(\Lambda|\vec{T}_{\vec{R}})\times \prod_{r = 1}^R P_{LC,i_{1_r}} \times \prod_{r = 1}^R P_{i_{L_r},RC}, \nonumber\\
\end{eqnarray}
where LC is the left censor (Jan. 1st, 2005) and RC is the right censor (Dec. 31st, 2009).

\subsubsection{Determining Cluster Membership}

In our algorithm we assign each patient to a profile based on the maximum posterior likelihood of the patient for each profile.  Let $\vec{Z}_{\vec{R}}$ be a latent variable vector $(\vec{Z}_1, \vec{Z}_2, \ldots, \vec{Z}_R)$, following a multinomial distribution and containing the latent profile membership of patient $r$, for $r \in \{1,\ldots,R\}$. Together the vectors $(\vec{X}_{\vec{R}},\vec{T}_{\vec{R}},\vec{Z}_{\vec{R}})$ form the complete data on the patient population under our model assumptions. However, because $\vec{Z}_{\vec{R}}$ is unknown, we must infer the $\vec{Z}_{r}$ from $\vec{X}_r$ and $\vec{T}_r$, specifically the posterior (conditional) likelihood $P(\vec{X}_r, \vec{T}_r|Z_{rk}=1)$, the probability that patient $r$ belongs to profile $k$ given $\vec{X}_r, \vec{T}_r$:
\begin{eqnarray}
P(Z_{rk}=1 | \vec{X}_r, \vec{T}_r)=\frac{P(\vec{X}_r, \vec{T}_r| Z_{rk=1})P(Z_{rk}=1)}{P(\vec{X}_r, \vec{T}_r)}\nonumber \\
 \propto P(\vec{X}_r, \vec{T}_r | Z_{rk}= 1).
\end{eqnarray}
Here, $P(\vec{X}_r, \vec{T}_r)$ will be constant for all $k$ and thus can be ignored.  Likewise, without any {\it a priori} knowledge of the system, we set $P(Z_{r1}=1)=P(Z_{r2}=1)=\cdots=P(Z_{rK}=1)$.  Therefore, profile membership will be solely determined by the posterior likelihood $P(\vec{X}, \vec{T}| Z_{rk} = 1)$. That is, each observation is assumed to belong to the profile which produces it with the greatest posterior likelihood.

\subsubsection{Model Selection}

We seek to find the optimal clustering of sequences, given by $\vec{Z}_{\vec{R}}$, such that the BIC score is maximized.  The BIC is an objective function that balances the tradeoff between maximizing the likelihood function while minimizing model size.  For a model $M$,
\begin{equation}
BIC(M) = \ell(M) + |M|\cdot\log(R)/2,
\end{equation}
where $\ell(M)$ is the log-likelihood of the model $M$, $|M|$ is the model size and $R$ is the number of patients.  Given the transition and interarrival parameters for the set of patients in profile $k$, $P_k$ and $\Lambda_k$, for $k \in  1,\ldots, K$, $\ell(M)$ is given by taking the log of the likelihood function, (5),
\begin{eqnarray}
\ell(M) = \sum_{k=1}^K    \sum_{r=1}^R  \ell(P_k|\vec{X}_{r}) + \ell(\Lambda_k|\vec{T}_{r})\nonumber\\
+ \sum_{r = 1}^R P_{LC,i_{r1}} + \sum_{r = 1}^R P_{i_{rL_r},RC}.
\end{eqnarray}
For model $M$ with $K$ profiles, we will estimate $K S (S+1)-1$ parameters for the transition matrices, $P_k, k \in \{1,\ldots,K\}$, and $K S^2$ in the interarrival matrices, $\Lambda_k, k \in \{1,\ldots,K\}$.


In a previous paper, the authors used an EM algorithm to perform model estimation.  However, such an algorithm requires the user to pre-specify the number of profiles, $K$, regardless of the number of true profiles.  Additionally, each initialization may produce a different outcome, implying that a global optimum is not necessarily reached with each clustering result. However, with a satisfactory initialization the output will be nearly optimal without complex calculation.

Other researchers favor a tree-based algorithm, where a distance metric is used to determine splits in the set of observation \cite{Ramoni:2002}.  Ramoni, et al., use the BIC in conjunction with the KL distance to perform agglomerative hierarchical clustering.  However, a top-down approach is warranted in our case since we can choose a reasonable stopping point in the algorithm where the smallest number of splits explain the predominant patterns in the system. In contrast with the EM algorithm, the benefit of such a tree-based algorithm is that the the number of clusters can be determined after the clustering analysis is performed. However, it may not be guaranteed to maximize posterior likelihood of cluster membership.  Therefore, we propose a joint tree-based, EM optimization algorithm that maximizes the BIC criterion.

\subsubsection{The Algorithm}

As $K$ and $R$ increase, it becomes computationally intractable to consider all possible partitions to find the maximum BIC score.  Therefore, we present an algorithm that searches for a nearly maximal BIC at each iteration.  Our algorithm, as in \cite{Ramoni:2002, Ramoni:2000}, is guided by the Kullback-Leibler (KL) distance:
\begin{equation}
KL(Q_1||Q_1) = \int Q_1(x) \log\left(Q_1(x)/Q_2(x)\right)dx,
\end{equation}
where $Q_1$ and $Q_2$ are the probability distributions under comparison.  Specifically, we find the KL distance between the transition distribution out of each of the $s_i$ for each individual sequence and the entire set of sequences in a given profile and then average across the $s_i, i \in \{1,\dots,S\}$. (We provide the derivation of the KL distance in Appendix B.) We then order the average KL distances and find a nearly optimal partition in the observations to use as the initialization of the EM algorithm to maximize the posterior likelihood function. An overview of the algorithm is given below:
\begin{enumerate}
\item We begin with the null assumption, $H_0$, that all patients in a set belong to one profile. Find the population MLEs, $\bar{\Lambda}_{ij}$, and the transition matrix $\bar{P}_{ij}$ under the null hypothesis.  Calculate the $BIC_0$ value.
\item Calculate the average KL distances between individual sequences and the one profile (null hypothesis), $D_{ave}(P,\Lambda|| \bar{P},\bar{\Lambda})$.
\item For a sufficiently large, equally-spaced set of the ordered average KL distances, (say, 50), $D_{(i)}$, let $W_{D_{(i)}}^-$ be the set of patients with average KL distances from the null distribution less than $D_{(i)}$, and $W_{D_{(i)}}^+$ be the set of patient with average KL distances from the null distribution greater than $D_{(i)}$.  For each partition, $\{W_{D_{(i)}}^-,W_{D_{(i)}}^+\}$ , calculate the $BIC_A$ corresponding to the $BIC$ value of the alternative hypothesis, $H_A$, that the set of sequences should be partitioned into two profiles.
\item Consider the partition $\{W^{*-}_{D_{(i)}},W^{*+}_{D_{(i)}}\}$, such that the $BIC$ score is maximized.  Let this partition be the initialization for the EM algorithm.  Recalculate the $BIC$ score, call it $BIC_A^*$ after the iterations of the EM algorithm.
\item If $BIC_A^* > BIC_0$, then divide the sequences into distinct profiles.  Repeat steps (1)-(4) until no more divisions are made.
\end{enumerate}

\subsection{Deriving Simple Utilization Profile Visualizations}

By employing stochastic models for clustering utilization sequences we can further derive stochastic provider networks via the transition matrices, allowing visualization of the utilization behaviors as networks across providers of different types. The primary inputs for the stochastic provider networks are the transition matrices. Specifically, the six provider types, CL, ER, HO, NP, PO, and RX are the nodes in a directed graph.  The directed edges represent transition probabilities between two provider types, for example, the transition of patients from the emergency room to a physician's office visit.  For a simplified representation, the networks only include nodes such that a total of 90\% volume is represented. we use different types of arcs for different levels of transition probabilities to better  identify nodes that are most visited within each profile.

\subsection{Assessing our Clustering Algorithm}

\begin{figure}[!t]
\begin{center}
\includegraphics[width=88mm,height = 60mm,angle=360]{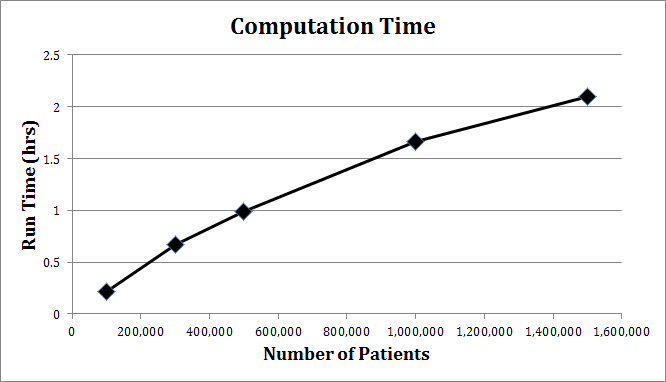}
\caption{\baselineskip=10pt The runtime in hours of a single iteration of our algorithm plotted against the number of simulated patients.}
\label{hilto1}
\end{center}
\end{figure}

We highlight five important properties of our clustering algorithm \cite{Andreopoulos:2009}:
\begin{itemize}
\item {\it Robustness}: Defined as the ability to detect outliers.  Our algorithm will place every observation within one profile, but as more divisions are made, the outlying observations become evident in low-membership profiles.
\item {\it Minimum user-specified input}: By combining the EM algorithm with a hierarchical framework we do not need predefine parameters such as the number of profiles in the algorithm.
\item {\it Scalability}: We simulated 5 different settings of $R$ patients ($R=100K$, $300K$, $500K$, $1M$, $1.5M$) and determined the run time of a single iteration of the algorithm. See Fig. \ref{hilto1} for results on the runtime of the algorithm.  In our study with over 100K patients in each state, the algorithm ran to completion through 8 iterations in approximately 3 hours.
\item {\it Computational complexity}: The primary computational steps involved in fitting a patient sequence to an MRP rely on simple counting and averaging, while the computation of posterior likelihood relies on multiplication.  All of these steps have computational complexity of order $O(n)$.  The sorting step of the posterior likelihoods is the most computationally expensive with order $O(n\log n)$. Therefore, the computational complexity of our algorithm is $O(n\log n)$.
\item {\it Visualization feasibility}: We translate the transition matrices into stochastic provider networks to produce simple visualizations of the utilization behaviors with each profile.  The ability to quickly digest information on the similarities and differences between the different stochastic provider networks is a major advantage over simply providing the resulting estimated transition matrices as it can play an integral role in decision support systems.  Moreover, we allow for different levels of visualization granularity of potentially complex healthcare systems. That is, the clusters of utilization behaviors can be split further into distinct profiles to reach a desirable balance between the number of profiles and intra-profile complexity.  This is especially important when there are potentially a large number of event types.
\end{itemize}

\section{Results}

In this section we summarize the results of our pilot study on pediatric asthma patients on Medicaid in GA and NC for the years 2005 through 2009.  We begin with 1.8  and 2.4 million total claims in Georgia (GA) and North Carolina (NC) for patients with a primary diagnosis of asthma which are translated into 754,597 and 1,224,579 visits for GA and NC, respectively.

\subsubsection{Graphical Representations}

\begin{figure*}
\begin{center}
\includegraphics[width=130mm,height = 80mm,angle=360]{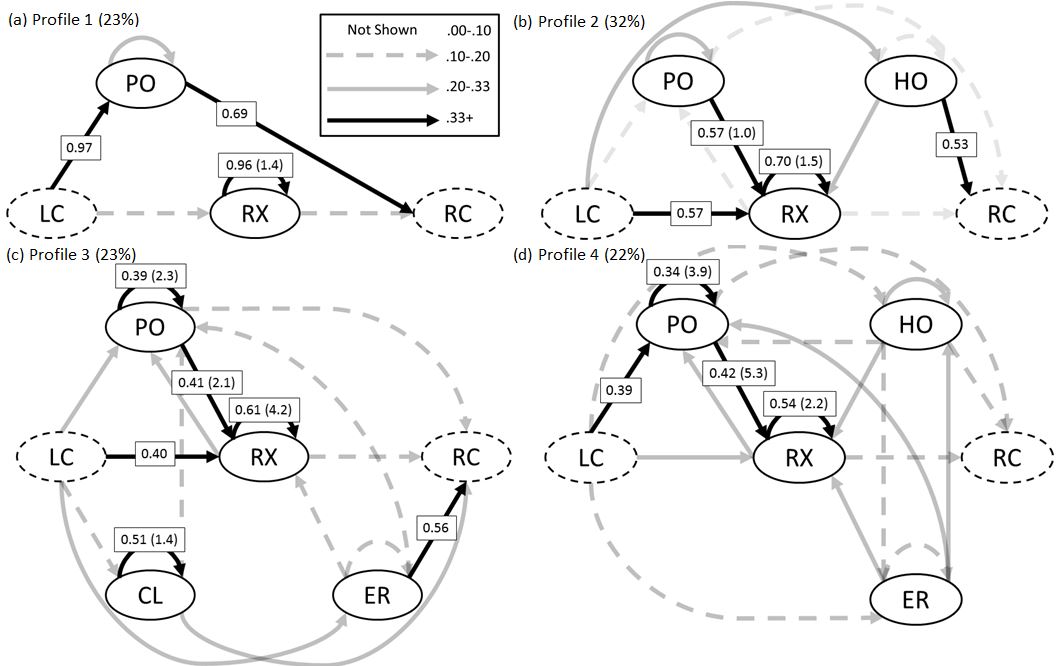}
\caption{\baselineskip=10pt Network graphs of estimated utilization profiles of GA. Transition probabilities are given on each edge along with the average interarrival times measured in months in parentheses.}
\label{hilto2}
\end{center}
\end{figure*}

\begin{figure*}
\begin{center}
\includegraphics[width=130mm,height = 80mm,angle=360]{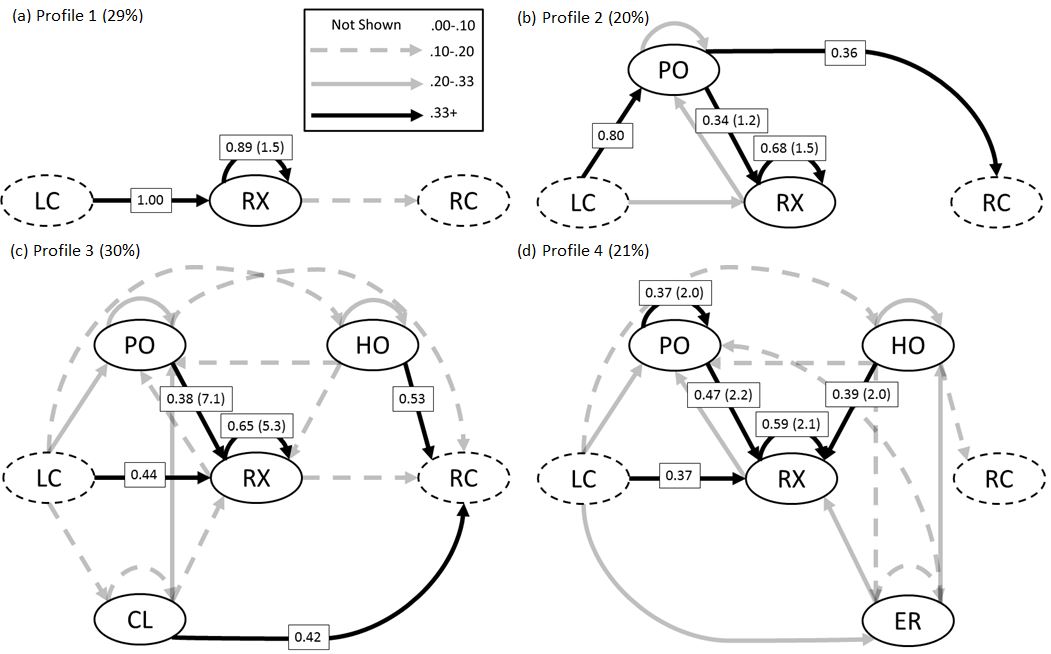}
\caption{\baselineskip=10pt Network graphs of estimated utilization profiles of NC. Transition probabilities are given on each edge along with the average interarrival times measured in months in parentheses.}
\label{hilto3}
\end{center}
\end{figure*}

Figures \ref{hilto2} and \ref{hilto3} are visual representations of the estimated utilization profiles as probabilistic network graphs.  We only include high-traffic nodes in these graphs, such that 90\% of the overall volume of encounters is summarized.  The nodes are labeled by the provider types corresponding to the contributing states of the utilization sequences in each underlying profile and edges are a visual representation of the the estimated transition probabilities between nodes or states. We provide the complete set of interarrival times between the different states in Tables II and III in Appendix D.  The legend describes our choice for visualization of the transitions between providers based on the transition probabilities between states. Across all networks, we include the LC (left censoring) and RC (right censoring) nodes specifying the beginning (2005) and the end (2009) year of the study.

\subsubsection{Utilization Networks for GA}

The network graphs of the four utilization profiles we highlight from GA are displayed in Figure \ref{hilto3}.  Our decision to highlight these four utilization profiles is described in Appendix C.

\underline{\it Profile 1}:  For patients in this profile, the initial probability of visiting PO is extremely high (0.97) while the probability of having repeat visits to PO is low, with average interarrival time of 8.2 months. Likewise, the initial probability of a RX prescription is low but the probability of repeat encounters is extremely high (0.96), with an average interarrival time of 1.4 months. There are no directed edges between PO and RX indicating that this profile consists of those patients who either visit PO or RX but not both, and would likely be divided into separate profiles in later iterations of the algorithm.

\underline{\it Profile 2}:  Patients in this profile have a high expected number of RX encounters, equal to 4.31, and PO visits, equal to 1.09, with a low expected number of HO, equal to 0.52.  There are many directed edges into RX with high probability (0.57 - PO $\rightarrow$ RX, 0.70 - repeated RX encounters), with no directed edges between PO and HO. The interarrival times into RX are also low (1.0 month for HO $\rightarrow$ RX, 1.1 months for PO $\rightarrow$ RX, 1.5 months for RX refills), and the interarrival time from HO to PO is 0.6 months. Although HO is present in this profile, repeat admissions into HO are infrequent, with an average interarrival time of 7.3 months.

\underline{\it Profile 3}: The expected number of visits  is 0.46 for CL visits, 0.5 for ER, 1.84 for PO and 3.30 for RX prescriptions, with many directed edges into PO and RX. The high number of RX prescriptions is due to many directed edges into RX from ER and PO as well as the high probability of repeat encounters with relatively high interarrival times compared to the other profiles of 4.2 months. PO likewise receives a high number of visits because of a large number of directed edges, although with low probability, from the other three provider types. Although ER is present in this profile, the readmission into the ER are infrequent, with 8.5 months on average between visits.

\underline{\it Profile 4}:  The expected number of visits to ER, HO, and PO are higher than in the previous profiles with 0.72, 1.13, 2.53, respectively, while RX still has a high number of encounters, equal to 4.03. The interarrival times between consecutive RX encounters are low on average at 2.2 months and interarrival times into ER and HO are overall high, with the lowest being HO $\rightarrow$ HO at 4.0 months.

RX is present in all four profiles, with high expected number of encounters in Profiles 2-4. The PO/RX relationship is highly prevalent, judging by the high transition probabilities between the two.

\subsubsection{Utilization Networks for NC}

The network graphs of the four utilization profiles we highlight from NC are displayed in Figure \ref{hilto3}.  Our decision to highlight these four utilization profiles is described in Appendix C.

\underline{\it Profile 1}: This profile consists of patients primarily on asthma-controlled medication, where the expected number of RX (re)fills is equal to 9.34 over the study period. The probability of RX refills is high at 0.89, while the interarrival time between consecutive RX encounters is low (1.5 months). These patients rarely visit physician offices (less that 10\% of the utilization in this profile and hence not present) and they almost never visit ER or have hospitalization. This group of patients could be used as a baseline to compare patients with  other utilization profiles.

\underline{\it Profile 2}: The expected number of RX encounters are lower in this profile (3.14) than Profile 1, with more expected visits to PO, equal to 2.35.  A strong connection between PO and RX is clear, with a stronger directed edge going from PO to RX, implying RX prescription fills after a physician office visit. The probability of RX refills is high, equal to 0.68, with a low average interarrival time of 1.5 months.   The average interarrival time between PO visits is higher (6.3 months).  Hence, patients in this profile tend to visit physician office more often than those in Profile 1, with insignificant ER utilization or hospitalizations.

\underline{\it Profile 3}: Patients in this profile have an overall lower number of visits to RX and PO (equal to 2.51 and 0.97, respectively), while CL and HO add more visits, with an expected number equal to 0.30 and 0.36, respectively.  This profile has many similarities to Profile 3 of GA, with much higher average interarrival times between RX encounters equal to 5.3 months, but also with high average interarrival times between HO readmissions equal to 5.8 months.  Transitions from HO to PO and to RX have high interarrival times at 9.9 months and 9.7 months, indicating non-adherence to follow-up treatment for controlling asthma.

\underline{\it Profile 4}: This profile primarily consists of RX and PO visits, with expected numbers equal to 7.52 and 4.16, respectively, while ER and HO add fewer, with expected numbers equal to 0.99 and 1.16, respectively.  There are strong connections between PO and RX, and many directed edges with high probability into RX.  Here the average interarrival times between consecutive readmissions to the ER and HO are 4.0 and 2.8, respectively, while the interarrival times between PO and RX encounters are lower, equal to 2.0 and 2.1, respectively.  All the interarrival times from ER and HO to PO and to RX are low ranging between 2.2 months and 2.5 months. Hence, patients in this profile display higher variation in their healthcare utilization for asthma than in the other three profiles.

\subsubsection{Comparing Utilization in GA \& NC}

\begin{figure*}
\begin{center}
\includegraphics[width=125mm,height = 80mm,angle=360]{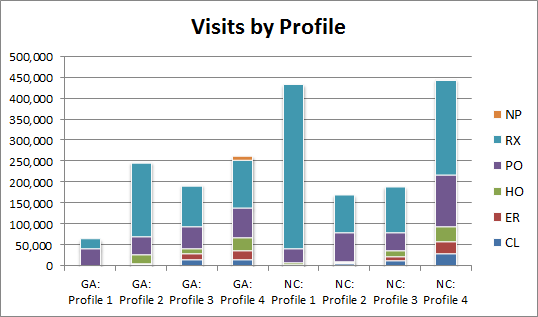}
\caption{\baselineskip=10pt A chart plotting the total number of visits to each provider type from all patients per profile during the years 2005 - 2009.}
\label{hilto4}
\end{center}
\end{figure*}

The network graphs for the two states show remarkable similarities between the longitudinal utilization profiles across both states; particularly, profiles Profile 1 of GA and Profile 2 of NC are similar as well as Profiles 3 and 4 of both GA and NC.  Other commonalities include the apparent prominent relationship between PO visits and subsequent RX encounters, with high probabilities, indicating well-managed asthma patients.  In all but Profile 1 of GA and NC there are directed edges between the two provider types, routinely with high probability and low average interarrival times.  Likewise, as shown in Figures \ref{hilto2} and \ref{hilto3}, there are no connections between PO or RX and HO or ER with transition probability greater than 0.33.  By examining the visits by provider type bar chart in Fig. \ref{hilto4}, we find that GA has more uniformity and variation between the provider types across the four profiles.  The major differences between the two states lie in the high concentration of RX visits in NC (67\% versus 54\% in GA), and the relatively high proportion of ER and HO visits in GA (13\% versus 8\% in NC).

\section{Conclusion}
In this paper we introduce a data science framework for extracting, analysing and integrating large, highly-sensitive claims information for deriving simple graphical interpretations of healthcare utilization. The objective is to characterize and visualize underlying profiles of patient-level utilization behaviors. Our framework begins with manipulation and processing of large flat files of administratively coded claims into meaningful data in the form of streamlined utilization sequences. The patient-level utilization sequences are then the input for a scalable model-based clustering analysis for discovering the underlying utilization profiles. Our methods are both rigorous and general, with applicability beyond the case study in this paper.

We pilot our study with Medicaid claims data across five years, 2005-2009. We extract data for only a subset of patients, particularly, asthma-diagnosed children older than 3, and we focus on two states, Georgia and North Carolina.


Our study emphasis is on healthcare utilization as it is at the core of critical aspects of healthcare delivery, including healthcare access, expenditure and cost, prevention and chronic disease management \cite{Rouse:2014}. We also focus on the Medicaid system as the test bed for our analysis because caring for the disadvantaged populations, particularly Medicaid children, is one priority of the current health policies in the United States, with potential impact on reducing health and healthcare disparities, and on containing the associated costs \cite{ACA:2010}. Medicaid constitutes the primary source of coverage for low-income children in the United States.

An important aspect of the Medicaid benefits system is that its implementation and reimbursement structure vary by state. Due to these state-based differences in the implementation, the effectiveness of the program also varies greatly by state. Thus, by comparing utilization of care across states one can reveal the impact of these variations on the care ecosystem.


Particularly, we chose Georgia and North Carolina for this comparison because the demographics of the pediatric populations are very similar (30-50\% minority population \cite{Census:2010} and approximately \$37,000 Per Capita Personal Income 2012 \cite{BBER:2013}), although they have different care-coordinated systems. While North Carolina has a state-coordinated Medicaid system, Georgia's Medicaid patients are primarily managed by three Medicaid Managed Care Organizations with a reasonably small percentage of children under the fee-for-service care practice \cite{Kaiser:2014}. According to the 2007 ranking of states based on the Medicaid eligibility, scope of services, quality of care, and reimbursement obtained by the Public Citizen Health Research Group \cite{Arellano:2007}, North Carolina is ranked in the second quartile and Georgia is ranked in the third quartile.


With similar Medicaid populations but different care coordination systems and effectiveness rankings, we find some striking similarities in the longitudinal (multi-year) utilization behaviors for pediatric asthma care. 

Both states have an underlying profile including patients primarily visiting a physician's office (Profile 1 in GA and Profile 2 in NC). Likewise Profile 2 in GA and  Profile 1 in NC have a high probability of filling a prescription for asthma-controlled medication (higher for North Carolina than for Georgia) but a lower probability of PO visits (lower for North Carolina than for Georgia). The transition probabilities are low connecting ER/HO to PO and vice versa, with stronger links between medication and physician offices, suggesting that the more variational clusters also include a proportion of patients that primarily visit physician office with sporadic (low probability) visits to the emergency department or with hospitalizations. This indicates that the majority of patients utilizing the physician's office in the variational profiles adhere to a great extent to evidence-based practices for asthma care.

A third noteworthy finding is the prevalence of clinic visits in Profile 3 for both states, where clinics refer to federally-qualified and rural health clinics. This is not surprising since Medicaid children rely heavily on care from clinics located in underserved areas. Importantly, patients with clinic visits have a higher probability to follow up with a phyician's office visit rather than visit ER or have an hospitalization for both states.

An important dissimilarity across the two states is the proportion of patients with regular PO visits (physician's office visits make up 59\% of non-RX (re)fills in Georgia, while they make up 64\% of non-RX (re)fills in North Carolina). Profiles 1 \& 2 in North Carolina contain almost 50\% of patients where those patients primarily utilize the physician's office along with RX encounters; in contrast, Profile 1 (21\%) and approximately half of Profile 2 (roughly 16\%) patients in Georgia utilize the physician's office almost exclusively. Hence, in aggregate, North Carolina has around 50\% more of the patient population than Georgia that visit physician's offices to the exclusion of other provider types on a regular basis.

In both states, the average interarrival time of RX fills is very similar averaging 1.5 months in Profiles 1 and 2 of both states, 2.2 months in Profile 4, with Profiles 3 having the longest average interarrival times of 4.2 months in GA and 5.3 months in NC. When comparing the graphical networks, we also find that physician's office visits and medication fills nodes are represented strongly for both North Carolina and Georgia across all four profiles. On the other hand, emergency department or hospitalization nodes appear to serve only as intermediary connections for both Georgia and North Carolina, with a stronger presence in Georgia.

This study has several limitations.  One shortcoming in using claims data to infer utilization is that while we seek to make inference on an entire subpopulation, we capture realized and not potential utilization of the system \cite{Rouse:2014, DeVoe:2011}. First, the MAX files only include claims that have been reimbursed. Second, not all Medicaid-eligible children are enrolled or they have intermittent enrollment. Moreover, there will be a percentage of Medicaid-enrolled children who are undiagnosed due primarily to lack of healthcare access.  Therefore, estimates on the healthcare utilization are likely to be to be biased, particularly for the Medicaid population, where certain subgroups have difficulty in maintaining Medicaid coverage or are susceptible to particularly disparate utilization \cite{DeVoe:2011, Piecoro:2001}.  We provide further analysis of the enrollment patterns of the Medicaid children in our study in Appendix E.

While our model and its estimation and selection methods are computationally attractive, they can be extended further for relaxing some of the underlying assumptions. First, we do not include the mean times until the first event and the mean times between the last event because they are biased estimates of complete lifetimes due to the censored nature of our data. In doing so, we are unable to completely determine the consistency with which patients visit providers. For instance, with unbiased estimates of the arrival to the first event it would be clear if a patient waits a long time between groups of consecutive visits or utilizes the system at a fairly homogeneous rate across the complete study time span. Furthermore, in order to produce simple visualizations and minimize computational costs we assume the interarrival times to be exponentially distributed, conditional on the visit type.  More importantly, it is likely that covariates including age, condition severity, comorbidities, enrollment status and access play a role in the frequency of the visits.  However, this method does not capture the potential effects of these covariates on utilization.

Despite these shortcomings, our model allows for reduction of high dimensional utilization data into a one-dimensional vector containing cluster memberships, thus providing the means for policy-makers to easily simulate or visualize healthcare utilization and further study explanatory variables that could explain the variations across patient-level utilization profiles.

Even though this study has several limitations, it has some important implications for health care providers and policy makers. Importantly, following the care practice recommendations, if a child visits the emergency department for asthma care then he/she needs to be referred back to primary care \cite{NIH:2007}. In both Georgia and North Carolina, the transition from emergency department or from hospitalization to physician's office varies across utilization profiles, with very low probability of physician's office follow-up visits for the patients using emergency department and hospitalization regularly. Those follow-up visits vary with the patient's profile, indicating that different interventions should be considered for each of the profile of patients. More importantly, in both states, patients who are visiting emergency department regularly for asthma care are few, with long periods of time between readmissions.

Asthma-controlled medication uptake is strongly connected with physician's office visits across three profiles, and in one profile where it is not, patients are regularly taking medication with no significant severe outcomes recorded. From the strength of the links between physician's office and medication (re)fills, and lack of connection of those two event types to the emergency department, those patients who visit a physician's office on a regular basis while staying on asthma-controlled medication are unlikely to have emergency department visits in both states. This finding provides evidence that asthma can be controlled with regular physician's office visits and medication, with the potential of eliminating costly emergency department visits.




%

\section*{Acknowledgment}

The authors are thankful to Matt Sanders and Paul Dietrich in assisting with data safeguards and access as well as with the information technology infrastructure. The authors are thankful to Dr. Julie Swann for the leadership in the protocol submission of the use of the MAX Medicaid claims data to the Centers of Medicare and Medicaid and in the Internal Review Board approval process.  We are also thankful to our reviewers who have helped us improve this document.  This research has been supported by a pilot grant from the Institute of People and Technology (IPaT) and Children's Healthcare of Atlanta and by a grant from the National Science Foundation (CMMI-0954283).




\bibliographystyle{plain}
\bibliography{Thesis}

\end{document}